# Potential of volatile organic compounds as markers of entrapped humans for use in urban search and rescue operations

Paweł Mochalski[*a], Karl Unterkofler[a,b], Gerald Teschl[c], and Anton Amann[*,a,d]

[a] Breath Research Institute of the University of Innsbruck, Rathausplatz 4, A-6850 Dornbirn, Austria

[b] Vorarlberg University of Applied Sciences, Hochschulstr. 1, A-6850 Dornbirn, Austria

[c] Faculty of Mathematics, University of Vienna, Oskar-Morgenstern-Platz 1, 1090 Wien, Austria

[d] Univ.-Clinic for Anesthesia and Intensive Care, Innsbruck Medical University, Anichstr, 35, A-6020 Innsbruck, Austria

[*] *corresponding author:* e-mail: pawel.mochalski@uibk.ac.at, tel: +43 -512-504-24636. fax: +43-512-504-6724636

**Abstract**

Volatile organic compounds emitted by a human body form a chemical signature capable of providing invaluable information on the physiological status of an individual and, thereby, could serve as signs-of-life for detecting victims after natural or man-made disasters. In this review a database of potential biomarkers of human presence was created on the basis of existing literature reports on volatiles in human breath, skin emanation, blood, and urine. Approximate fluxes of these species from the human body were estimated and used to predict their concentrations in the vicinity of victims. The proposed markers were classified into groups of different potential for victim detection. The major classification discriminants were the capability of detection by portable, real-time analytical instruments and background levels in urban environment. The data summarized in this review are intended to assist studies on the detection of humans via chemical analysis and accelerate investigations in this area of knowledge.

## 1. Introduction

Earthquakes belong to the most frequent and catastrophic natural disasters affecting mankind. In the last century earthquakes occurred with an annual worldwide incidence of one million events (two earthquakes per minute) [1] causing more than 1.5 million deaths and affecting another 2 billion people [2]. Bearing in mind the increase of global urbanization and the fact that the most populous cities are located in seismic zones, it is reasonable to assume that these numbers will rise considerably in the nearest future [3]. In contrast to many other disasters earthquakes not only cause many deaths, but also many traumatic injuries and massive entrapment of survivors in collapsed buildings [1, 3, 4]. While about 50% of survivors are found and rescued quickly by bystanders, or other civilians [5, 6], the remaining ones are subjected to prolonged entrapment under complex debris. Their extrication frequently requires trained and specially equipped rescuers. Since the survivability of victims is directly related to the entrapment time [1], the early location of entrapped victims is of utmost importance for urban search and rescue (USaR) operations. Until now, a number of technical tools have been employed to reduce the length of entrapment. These embrace, e.g., fiber optic cameras (borescopes), acoustic probes aiming at voices, or heartbeats, thermal cameras, and sonars [5]. Nevertheless, search-and-rescue (SAR) dogs remain indispensable for rescue teams and are commonly recognized as the golden standard in this context [7]. Search dogs exhibit excellent scenting skills, are able to search relatively large areas in a short period of time and can work in areas that are deemed unsafe, or inaccessible to human rescuers. They, however, exhibit a number of limitations. Their working time is relatively short and restricted to approximately 30 min (with a subsequent break of 2 hours) and their training is time-consuming and expensive. Moreover, they respond poorly being stressed or frustrated and can easily be injured in highly toxic and harsh disaster environment [8]. All these constraints caused a huge demand for novel detecting tools, which could complement, or even replace search dogs during USaR operations. The fact that SAR dogs can detect survivors in highly contaminated disaster sites implies that there is a human-specific chemical signature in void spaces of collapsed buildings and that the analysis of this signature could be a valuable detection tool. Unexpectedly, this approach has received little attention and was limited to carbon dioxide sensing [9]. This is surprising as small molecule volatile species are often the final products of vital metabolic pathways occurring in human organism and could therefore serve as signs of live in the context of rescue operations [10-12]. Indeed, there is growing evidence provided by a number of very recent but early studies suggesting that some constituents of the human scent could be employed for this purpose and thereby considerably improve the effectiveness of rescue teams [13-16]. Apart from the detection of victims, chemical analysis could provide the rescuers with the capability to recognize exposures to potentially toxic agents which can be present at disaster sites [17, 18]. Consequently, toxicological hazards and risks for humans and animals could be considerably minimized during rescue operations. Thus, in the context of USaR operations chemical analysis towards volatiles can be considered as a very promising field, which is, however, still in its infancy.

The primary goal of this review was the creation of a database containing constituents of the human scent having potential to serve as signs-of-life during USaR operations. The database was built on the basis of existing literature reports on volatiles in breath, blood, urine and skin emanations. It should be stressed here that only quantitative data were taken into consideration. In particular, by this we intended to provide a list of preliminary markers of human presence to be verified and complemented during future field studies. An effort was also made to estimate the approximate emission rates of these compounds from the human body as paramount factors determining their levels in the vicinity of survivors. A secondary goal was to predict the tentative levels of the preselected markers in void spaces of collapsed buildings and assess the capabilities of their detection by selected portable field analytical instruments against the urban environmental background.

## 2. Sources of human scent during entrapment

Volatile species forming the human scent during entrapment can stem from different biological fluids (breath, urine, blood, sweat) and organs (skin, lungs, bowels). Generally, sources of human-related volatiles can be classified into continuous and temporal ones. The former group embracing breath and skin emanations is particularly important in the context of victim detection, as it offers a long-lasting emission of potential markers of human presence. Moreover, breath holds here a distinguished status since the breath-borne volatile species can help to differentiate between living and dead victims.

Temporal sources such as blood or urine have a more transient contribution to human scent; nevertheless, this impulse-type contribution cannot be neglected. The occurrence of this impulse of volatiles is difficult to predict; however, it is reasonable to assume that emission of blood-borne species should appear at the early stage of entrapment as a result of injuries induced by the disaster. Furthermore, urine- and blood-borne compounds are expected to strengthen the location signal provided by breath markers of human presence due to the physiological dependencies between these fluids. On the other hand blood and urine should be considered as limited reservoirs of species tending to dry out and/or clot.

The emission rates of volatiles from the aforementioned sources depend on the physiological and medical status of the victim (injuries, dehydration, shock, diet, history of environmental exposure, drug intake, etc.), conditions in the entrapment scene (confined space volume, type of collapse, temperature, humidity, oxygen content), and the time of entrapment. In particular the disaster event and the entrapment induce a number of neuroendocrine, metabolic and physical responses [19]. These can comprise, e.g. intense emotional stress, physical shock, hypermetabolism (manifested by hyperglycemia, hyperlactatemia, and protein catabolism), immunological responses, and up-regulation of hormones' secretion. All these factors inevitably influence the production and emission of volatiles by a human organism. Unfortunately, this impact is poorly understood. This is due to the limited quantitative data on the emission rates of VOCs from the human body, limited knowledge of human physiology during entrapment as well as ethical and methodological problems related to the simulation of entrapment under laboratory conditions. As a consequence, the emission of volatiles from entrapped individuals and their propagation during entrapment are very difficult to estimate. In this context emission rates of volatile species from healthy volunteers at normal conditions seem to be the only reasonable surrogate of these parameters. Moreover, the understanding of the production and initial composition of the human-specific chemical signature is of particular importance for modeling the behavior of potential markers of human presence in the surroundings of the entrapped person and determines the selection of on-site, real-time, and handheld analytical instruments, which could be used for the field detection of entrapped victims.

One of the main goals of this work was to pre-select potential markers of human presence and to estimate their emission rates from the human body on the basis on existing literature data on volatile organic and inorganic compounds in breath, urine, blood, and skin emanations. Several prerequisites have been assumed to achieve this goal. First, only emissions via breath and skin were used to calculate the total fluxes of volatiles from the human body. This stems from the fact that the occurrence and intensity of urine-, or blood-borne VOCs is much more variable and difficult to predict. Second, only omnipresent and reliably identified compounds were used to construct the set of potential markers of human presence. Here, a compound was recognized as omnipresent when it was reported to have an incidence of at least 80%. The threshold of 80% was arbitrarily chosen. The reliable identification was defined as the identification that is based on several methods and thereby providing unequivocal results. For instance, in case of GC-MS studies compounds identified exclusively on the basis of a spectral library match (e.g., NIST) without taking into account the retention time (or retention index) were excluded as only tentatively identified. Finally, only species having clearly higher levels in breath than in room air were recognized as produced by the human body and thereby contributing to the formation of human scent. It should also be stressed here that compounds have not been pre-selected with respect to their origin as it still has not been elucidated in sufficient depth and in many cases is a matter in dispute. Table 1 lists volatile organic and inorganic compounds, which fulfilled the aforementioned requirements. An effort was made to provide for each compound data from different literature sources and obtained by different analytical techniques to improve the reliability of calculated fluxes.

### 2.1. Breath

Exhaled breath contains a wide range of volatile compounds capable of providing invaluable information on normal and disease processes occurring in an individual as well as his/her environmental exposure to pollutants/toxins, or microorganisms' activity in the body [10-12]. Its attractiveness in biomedical applications stems from the fact that it is readily and noninvasively obtainable and may be sampled as often as it is desirable without discomfort for a subject. Moreover, concentration levels of breath compounds can respond rapidly to changes in human physiology and thereby provide near real-time information on processes occurring in the organism [20-23]. In the context of urban search and rescue operations breath volatiles play a fundamental role as breathing can be considered as a sign of live and the breath-specific species can help to distinguish living victims from dead ones. Due to the aforementioned reasons breath volatiles received enormous attention in the literature. Moreover, the majority of published clinical studies provides also data obtained for control populations (healthy volunteers, hospital personnel, etc.), which potentially could be useful for the purposes of this work. Unfortunately, a considerable fraction of the existing sources suffers from several disadvantages such as reporting of only qualitative or semi-qualitative data (e.g., peak areas, relative abundances), absence of detection frequencies of observed species, or absence of room air (inhaled air) data. Consequently, their value for the goals of this work is limited. Moreover, the literature sources have been additionally constrained to the ones providing data for the end-tidal exhalation phase and mean concentrations of species under scrutiny. Such an approach aimed at the reduction of the variability of results induced by different sampling protocols.

Table 1 lists 34 breath volatiles which were selected using the aforementioned criteria together with their literature levels in the end-tidal exhalation segment. These concentration data were used to calculate the breath fluxes of compounds of interest. First, for each compound a weighted arithmetic mean of means provided by all considered literature sources was calculated. The weight factor was the population involved in the particular study. Next, these means were converted into nmol×L$^{-1}$. Finally, the emission rates expressed in nmol×min$^{-1}$×person$^{-1}$ were calculated assuming an alveolar ventilation of 3.3 L×min$^{-1}$, which is typical for sleep [24]. Such an approach stems from the fact that entrapped victims are frequently unconscious, or drift between sleep and consciousness over the course of entrapment [5]. Since the real values of alveolar ventilation during entrapment are difficult to predict and can be considerably affected by the conditions in the entrapment environment, sleep seems to be a good (although simplified) surrogate model in this context. The calculated breath fluxes of compounds of interest are presented in Table 1 and Figure 1. With the exception of $CO_2$ the estimated emission rates range from 0.03 to 524 nmol×min$^{-1}$×person$^{-1}$. Within this group the highest values were for CO (524 nmol×min$^{-1}$×person$^{-1}$), ammonia (91 nmol×min$^{-1}$×person$^{-1}$), acetone (60 nmol×min$^{-1}$×person$^{-1}$), and methanol (45 nmol×min$^{-1}$×person$^{-1}$). The majority of compounds (56%) exhibited breath fluxes falling below 1 nmol×min$^{-1}$×person$^{-1}$ (considering means).

### 2.2. Skin

Skin, next to breath, is a principal source of human scent constituents, as it offers a long-lasting emission of VOCs from a relatively large area. The composition of skin emanation in humans has received considerable attention and numerous reports dealing with this issue can be found in the literature [25-29]. Although these studies reported a large number of species, the majority of them yield only qualitative data, i.e. names of identified compounds and possibly their occurrence in skin emanations. Moreover, the GC-MS-based studies provide mainly tentative identification of these species based on peak spectra that were checked against commercial mass spectral libraries (e.g., NIST). Quantitative data (emission rates) are relatively sparse [29-35] and usually determined for peripheral skin (hand, arm, or leg). Such a sampling protocol is obviously convenient for human subjects; however, the obtained results are not necessarily representative for the remaining parts of the skin. This stems from the fact that due to the differences in the distribution of sebaceous glands, the composition and thickness of human sebum vary between different parts of the body [28, 36] and the emission of volatiles can reflect these variations. The whole body emission data are even sparser, although in the context of this review the most valuable ones [32]. Thus, the assessment of the contribution of skin-borne species to the formation of a human-specific chemical fingerprint may suffer from the shortage of reliable data.

The skin fluxes of compounds of interest from the whole human body were estimated in several steps. First, the emission rates reported for a certain skin area (e.g., $cm^2$) were rescaled to the total skin area of the volunteer. This was done using the volunteer's skin area estimated by the formula given by Mosteller [37], or (in the case of the unavailability of the volunteers' data) by taking the average area of the human skin of 1.7 $m^2$. Next, the emission rates were converted into nmol×$L^{-1}$×$person^{-1}$ and the weighed arithmetic mean of means of all considered literature data was calculated. The weight factor was the population of the particular study. It should be stressed here that the use of peripheral skin data for the aforementioned purposes implies the underestimation of the calculated whole body fluxes. The obtained emission rates of potential markers of human presence from skin are presented in Table 1 and Figure 2. Their values vary from 26.5 µmol×$min^{-1}$×$person^{-1}$ to 0.02 nmol×$min^{-1}$×$person^{-1}$ for $CO_2$ and 3-methylfuran, respectively. Of 38 species only 5 ($CO_2$, ammonia, acetone, acetaldehyde, and 6-methyl-5-hepten-2-one) exhibit fluxes exceeding 2 nmol×$min^{-1}$×$person^{-1}$.

### 2.3. Urine

Urine is an important reservoir of human scent constituents. Until now, more than 230 volatile organic compounds belonging to different chemical classes (e.g., aldehydes, ketones, furans, pyrroles, terpenes, sulphur-containing compounds) have been identified in human urine [16, 38-41]. This high abundance of species results from the kidneys' pre-concentration capabilities. Thus urine in a certain sense offers an insight into the composition of blood volatile compounds. Nevertheless, in the context of USaR operations this source of markers suffers from several disadvantages such as unpredictable and temporal occurrence, or limited capacity. Its contribution to the total flux of volatiles from human body during entrapment is also difficult to estimate. It is reasonable to assume that the fluxes of some species (e.g., ketones) can be temporarily strengthened [14, 16] after the urinating event, however, this increase of emission will depend on their physicochemical properties. For instance, compounds well soluble in urine will be released much longer than poorly soluble ones [14]. Moreover, the entrapment conditions such as temperature, humidity, or dehydration can considerably affect the urination cycles and quantities. For the aforementioned reasons the contribution of urine-borne species to the total VOCs flux have not been assessed quantitatively within this review. Instead, we report only their omnipresence to indicate that this source can raise the total signal.

### 2.4 Blood

Apart from many deaths earthquakes typically result in many traumatic injuries. These injuries are highly mechanical and often multiple. The musculoskeletal injuries typically embrace lacerations, fractures, crush injuries, soft-tissue contusions, or chest trauma [1]. Consequently, the probability of victim's bleeding after the disaster event is relatively high. Although, the levels of volatiles in blood are generally lower than in urine (acetone is here an exception worth to be mentioned) the emission of blood species is much more predictable and should occur at the early stage of entrapment. Blood, however, as a source of volatiles shares the urine limitations. It is temporal, of limited capacity, and its contribution into the chemical signature is variable and depends on the victim's medical status. Thus, as in the case of urine we indicate in Table 1 only the omnipresence of blood species to stress the possible contribution of this source.

## 3. Potential markers of human presence

Altogether 47 compounds were selected as potential markers of buried victims (see Table 1) on the basis of skin and breath VOCs' data. Their total mean emission rates from the human body are given in Table 1 (column I) and shown in Figure 3. The tentative origins of these species in human organisms have been listed in Table 1 (column F). Within this set of species the most numerous chemical classes are aldehydes (23%) and hydrocarbons (21%). Other well-represented families are ketones (13%) and inorganic compounds (9%). It must be stressed here that this list should be considered as an initial library to be complemented and verified under field conditions rather than a closed and complete set of potential markers. In particular a number of omnipresent breath species were excluded due to the shortage of quantitative data, or contradictory information on their origin. The chemical pattern depicted in Figure 3 demonstrates that both permanent sources of volatiles in humans contribute considerably to the formation of a human-specific chemical fingerprint and numerous species stem from both breath and skin. This is not surprising as endogenously produced compounds can be distributed actively (vascular system) or passively (diffusion) among tissues and organs and finally released via breath, skin, or urine.

Four inorganic compounds ($CO_2$, CO, NO, and ammonia) have been preselected within this work. With the exception of NO they exhibit very high emission rates from the human body and thereby offer an enhanced possibility for the detection of victims in voids of collapsed buildings. They are predominantly released via breath. Only ammonia has a considerable skin emission component. Carbon dioxide appears to be the most natural candidate as a sign-of-life. It is produced endogenously in huge amounts and released almost exclusively via breath. Its total flux is approximately four orders of magnitude higher than the flux of the second most abundant compound – ammonia. Moreover, it is relatively inert and rapidly transportable by air currents; however, its levels can be influenced by high humidity and water absorption on debris materials [15], or fires in the voids of collapsed buildings.

Indeed, $CO_2$ has already been employed by shipping companies for stowaways detection in, e.g. harbor or airport locations [9]. Despite these advantages $CO_2$ poses some hazards for an entrapped victim. In case of the absence of air currents within the void spaces its levels will rise and the levels of oxygen will decrease leading to the victims' asphyxiation and death. Conversely, ammonia - another abundant inorganic compound seems to be released mainly through skin emanations. More than 85% of its total flux stems from this source. In healthy individuals ammonia is produced in the gut during bacterial breakdown of proteins [42]. However, analogous protein breakdown in the oral cavity or on skin surface may also contribute to the flux of this volatile [43]. It is unclear why ammonia exhibits such a considerable skin component. Perhaps skin emission is promoted by the rapid diffusion of $NH_3$ via tissues resulting from the low molecular mass of this volatile. The usefulness of ammonia as an indicator of human presence was suggested by several authors [13, 15]. Interestingly, the flux of this compound seems to differ between sleep and consciousness [15]. Since other species can share this phenomenon, studies on breath VOCs during sleep are of particular importance for USaR operations [44]. The production of CO in humans is ascribed to the endogenous metabolism of heme [45]. Relatively high levels in breath (at low ppm concentration) render this compound as one of the most abundant volatiles amongst species released by humans.

A total number of 6 ketones were found to be omnipresent in human breath and/or skin emanations and thereby valuable for the detection of humans. These were acetone, 2-butanone, 2-pentanone, 3-buten-2-one, 2,3-butanedione, and 6-methyl-5-hepten-2-one. Acetone is the major ketone produced in the human organism exhibiting high abundances in breath [46-48], blood [49], and urine [16, 38]. Several sources of this compound in humans can be indicated. These are (i) endogenous decarboxylation of Acetyl–CoA [47, 50], (ii) oxidative degradation of squalene on human skin [51], (iii) 2-propanol metabolism [52], and (iv) diet. However, the latter two are of minor importance. The high emission rate of 85 $nmol \times min^{-1} \times person^{-1}$, endogenous production and high volatility render acetone a very promising marker of buried victims. Indeed, early experiments aiming at the identification of markers of human presence indicated acetone as a compound having a great potential in this context [13, 15]. The remaining ketones were characterized by much lower emission rates ranging from 0.47 to 3.7 $nmol \times min^{-1} \times person^{-1}$. The origin of these species remains still ambiguous and therefore it is difficult to assess their usefulness during USaR operations. For instance, 2,3 butanedione appears to originate from butter consumption; whereas, 3-buten-2-one may be a product of isoprene degradation. 6-methyl-5-hepten-2-one is an interesting ketone released from human skin in considerable amounts. It stems from the oxidative degradation of squalene – a major component of human sebum. Squalene is a particularly interesting component of sebum as its levels are very low in other organs but particularly high in human skin and range from 12 to 20% of total skin surface lipids [53]. This high abundance is also unique to human skin as compared to other animals. Squalene is believed to be a natural antioxidant capable of neutralizing reactive oxygen species (ROS) [36, 54]. While exposed to ROS it degrades producing a wide range of semi-volatile and volatile products [36, 55, 56]. Thus, some of these compounds can be human-specific and thereby very valuable for USaR operations. 6-methyl-5-hepten-2-one could be a prototypic representative of this group. Although it was found to be emitted by some plants [51], in urban environments it could be a biomarker of human presence.

Aldehydes with 11 representatives were the most numerous chemical family amongst the compounds of interest. Interestingly, they predominantly originate from skin. With the exception of acetaldehyde, 2-propenal, n-hexanal, and n-propanal their breath fluxes are very small, frequently negligible as compared to the skin ones. This ample presence of aldehydes in skin emanation mirrors the oxidative stress inducing peroxidation of unsaturated fatty acids on skin. Apart from squalene sebum contains numerous long-chain fatty acids (up to 26 carbon atoms), linear or branched, predominantly saturated or mono-unsaturated [36, 57]. Strikingly, fatty acids fraction embraces very unique components like branched-chain species, or lipids with unique patterns of unsaturation [57, 58]. For instance, two predominant sebum lipids, sapienic and sebaleic acid have not been identified in other human tissues and in the sebaceous gland secretions of other animals [53]. Oxidative stress on skin surface causes peroxidation of these fatty acids with subsequent formation of numerous volatile products (aldehydes, ketones, hydrocarbons, alkohols, or esters) [59-62]. They are generated via β-scission of alkoxy radicals formed by the homolytic cleavage of fatty acids hydroperoxides. For example, oxidation of oleic acid leads to the release of n-octanal, n-nonanal, n-decanal, n-heptane, and n-octane [59-62]. Interestingly, some aldehydes are more abundant in skin emanations of older subjects (e.g., n-nonanal [28], or 2-nonenal_[60]) indicating some age-related changes in the fraction of skin fatty acids. Bearing in mind the multitude and diversity of fatty acids building the human sebum it is not surprising that skin emanation contains numerous members of this chemical family. Oxidation of fatty acids is not, however, the only source of aldehydes in the human organism. They can also be products of the endogenous oxidation of primary alcohols catalyzed by alcohol dehydrogenases (ADHs) [63], or stem from dietary sources. Thus, breath acetaldehyde can mirror ethanol metabolism, whereas, n-propanal may reflect the exposure to 1-propanol.

Regarding hydrocarbons, 10 species were preselected. Amongst them there are four alkenes, three alkanes, and three diens. Although hydrocarbons emitted by the human body have received special attention as non-invasive markers of numerous diseases or metabolic disorders [10-12] their sources in humans have not been elucidated in sufficient depth. Nevertheless, several metabolic pathways leading to the formation of hydrocarbons of interest can be indicated. A wide range of hydrocarbons (both saturated and unsaturated) is generated during cutaneous oxidation of the sebum components such as fatty acids and squalene. This mechanism is identical to the one responsible for the formation of aldehydes and proceeds as described above. Thus, n-octane was found to be the product of the oleic acid degradation [59], whereas, 2-methyl-2-pentene was reported to stem from decomposition of squalene [55]. Isoprene is an unsaturated hydrocarbon produced in humans in large quantities [64].

According to the current theory it is formed from isopentenyl pyrophosphate (IPP) and its isomer dimethylallyl pyrophosphate (DMAPP) in the mevalonic acid (MVA) pathway [64]. In animals and humans it has been suggested to be produced non-enzymatically by acid-catalyzed formation from DMAPP occurring in the cytosol of hepatocytes. Nevertheless, there is growing evidence provided by a number of recent studies suggesting that other endogenous metabolic sources may contribute to isoprene formation in humans [65, 66].

All preselected alcohols (ethanol, methanol, and 2-propanol) exhibit high abundances in the human scent. Their total emission rates range from 12.8 nmol×min$^{-1}$×person$^{-1}$ for 2-propanol to 45.1 nmol×min$^{-1}$×person$^{-1}$ for methanol. It is worth mentioning here that these species are also emitted via skin emanations in considerable amounts [67]. However, due to the shortage of quantitative data it is difficult to assess their skin-borne component in the total flux. Several sources of these compounds can be listed in humans. First, they can stem from dietary sources (e.g., fruits consumption) [68]. Methanol and ethanol can be produced by the bacterial flora in gut and/or oral cavity [69, 70] and 2-propanol can be the product of acetone metabolism catalyzed by alcohol dehydrogenases (ADHs) [71-73].

Amongst sulfur containing compounds there were two diet-related species (allyl methyl sulfide and methyl propyl sulfide) and dimethyl sulfide. The presence of allyl methyl sulfide in human tissues and fluids is attributed to the garlic consumption [74]; whereas, methyl propyl sulfide was shown to appear in human breath after onion intake [75]. Dimethyl sulfide (DMS) is a volatile reported to be omnipresent in human breath and blood [46]. Its production is ascribed to the metabolization of sulfur-containing amino acids methionine and cysteine in the transamination pathway [76]. Thus, in liver thiol S-methyltransferase forms DMS via the methylation of methyl mercaptane [76, 77].

Of terpenes DL-limonene exhibits the highest emission rate of 0.77 nmol×min$^{-1}$×person$^{-1}$. The fluxes of the remaining species from this chemical family (p-cymene and α-pinene) are notably lower and do not exceed 0.1 nmol×min$^{-1}$×person$^{-1}$. Despite their omnipresence the origin of these species in humans remains ambiguous. Nevertheless, diet seems to be the most probable reason of their occurrence [68, 78].

Amongst the remaining compounds there are three furans (2-methylfuran, 3-methylfuran, and 2-pentylfuran), acetonitrile, γ-butyrolactone, methyl acetate, and dimethylselenide. Several compounds from this set can stem from the metabolism of microbiota inhabiting the surface of human skin. For instance, fungi of the genus *Malassezia* naturally found on the human skin were found to produce an odor exhibiting high abundance of a homologous series of γ-lactones (C8-C12) at the presence of oleic acid or human sebum [79]. Although γ-butyrolactone could not be measured in the above-cited study due to some analytical limitations it appears plausible that also this compound could be a marker of fungi from this genus. Moreover, 2-pentylfuran was demonstrated to be produced by *Fusarium sp.* and *Aspergillus flavus* [80]; whereas, 3-methylfuran was found to be released by *Penicillium sp.* and *Aspergillus flavus* [81]. All these fungi species inhabit human skin [82]. Alternatively, 3-methylfuran might be produced also endogenously during the alkoxy radical-induced degradation of isoprene, as it was established in atmospheric studies [83, 84]. If so, isoprene could be considered as a ROS scavenger protecting skin surface from oxidative stress-induced damages. Nevertheless, additional experiments are necessary to pinpoint the role of isoprene in human physiology. Interestingly, acetonitrile a compound commonly being attributed to smoking habits [85] appears also in the scent of non-smokers [86, 87]. Perhaps additional endogenous sources contribute to the pool of this species in humans.

## 4. Changes of VOCs emission rates during entrapment

Entrapment conditions and disaster-related injuries notably affect physiology and biochemistry of humans [19]. Earthquakes typically cause highly mechanical and often multiple injuries. Amongst them musculoskeletal injuries such as lacerations, fractures, crush injuries, or spinal trauma are the most common ones [1]. These traumas induce numerous systemic complications and a number of neuroendocrine, metabolic, and physical responses. The latters can comprise, e.g. intense emotional stress, physical shock, hypermetabolism (manifested by hyperglycemia, hyperlactatemia, and protein catabolism), immunological responses, or up-regulation of hormones secretion [19]. Moreover, prolonged entrapment introduces additional complications such as, e.g. dehydration, starvation, or asphyxiation. Although the epidemiology of disaster-related injuries and complications received considerable attention [1], little is known how these conditions affect the production and emission of volatiles from the human body in general and markers of human presence in particular. This is due to the limited quantitative information on the emission rates of VOCs from human body, limited knowledge of their origin, and fate as well as ethical and methodological problems related to the simulation of entrapment in a laboratory environment. Nevertheless, despite these constraints several possible responses of the human-specific chemical fingerprint to the entrapment can be indicated.

Starvation, stress, and hypermetabolism (hyperglycemia) induce ketone bodies' production, which should be manifested by increased emission rates of acetone [19]. Abnormally high concentrations of acetone (ketoacidosis) in turn foster biotransformation of this compound into 2-propanol catalyzed by alcohol dehydrogenases (ADHs) and thereby increase the levels of this metabolite in the human organism [71-73]. Crush injuries are usually associated with traumatic rhabdomyolysis and release of products of muscle degradation. These compounds may include myoglobin, uric acid, potassium, lactic acid, or creatine kinase [1]. In excess, these species have toxic effects on distant organs. In particular high levels of myoglobin accompanied by acidosis and hypovolemia obstruct kidneys tubular flow and induce acute kidney injury. Impaired kidneys fail to eliminate the urea and precipitate to the rise of ammonia levels in tissues and fluids (hyperammonemia) [19].

The entrapped victim is inherently cut off from the predominant factors inducing oxidative stress on the skin surface such as UV radiation, or $O_3$. Thus, it can be expected that the skin production of oxidative stress-related species will be reduced shortly after entrapment. In particular this can affect emission rates of numerous aldehydes and hydrocarbons (see Table 2). Consequently, the applicability of compounds from this group may be limited to the initial period of rescue operations.

Diet contributes enormously to the pool of VOCs in the human organism. Myriads of volatile compounds are consumed as flavor constituents of food or beverages [68]. Some of them are of natural origin, whereas the others stem from human or bacterial food metabolism. These volatiles are next distributed amongst tissues and excreted via breath, skin, or urine. A number of compounds being potential markers of human presence can at least partly originate from this source, as shown in Table 2. Prolonged starvation will reduce the abundance of diet-related compounds in the human specific chemical pattern. Thus, they can have limited applicability during longer rescue operations. In particular, this problem can concern some well abundant species such as methanol, ethanol, or ammonia (produced mainly by bacteria in gut and/or oral cavity). Thus, the knowledge of the origin and metabolic fate of potential markers of human presence is of utmost importance for their verification.

## 5. Human-specific chemical signature at the entrapment site

Once emitted, volatiles forming the human scent are spread by air currents throughout the void spaces of collapsed buildings, interact with debris materials, and mix with environmental and disaster-related contaminants and/or toxic agents. All these factors can considerably distort the original human-specific chemical fingerprint. The type of the building construction and construction materials are considered to be critical factors in mortality and epidemiology of earthquake-related injuries. Survivors are frequently found in confined spaces of collapsed building structures. The retention of these voids depends on the collapse mechanism (e.g., pancake, lean-to, V-shaped [5]) and is much more likable in well-constructed, reinforced concrete, steel frame buildings than in masonry, brick, or adobe constructions [6]. The presence of void spaces as well as the type of debris materials will also affect the dispersion and air levels of potential volatile markers of human presence. Here, a very important parameter is the debris surface-to-volume ratio (SA:V) as it governs the surface chemistry. High values of SA:V favor the adsorption of volatile species on building materials and decrease their concentrations in void spaces. Moreover, these loses can be additionally boosted by the presence of dust and powdered building materials covering the rubble and buried victims. In particular, dust can notably suppress the emission of skin-borne species and thereby limit their applicability during USaR operations. Additional factors affecting the VOCs' levels in void spaces are temperature and especially humidity. Relative humidity over 90% induces condensation and formation of water films and thereby triggers wet chemistry. Thus, the knowledge of the surface chemistry of building materials can determine the applicability of human presence markers.

The interactions of VOCs forming the human scent with debris materials have already received some attention. Several authors investigated the permeation of urine-borne volatiles through layers of different building materials such as concrete, brick, or quartz stone [16, 88, 89]. Volatiles in the urine headspace were found to exhibit a concentration profile with an initial peak related to urinating, which next was washed out by the prevailing air currents. The influence of debris materials on these profiles depended on their fundamental physicochemical properties. Brick was demonstrated to be a much less adsorptive material for urine-borne species than concrete. Although concrete considerably reduced the observed levels of compounds, it prolonged VOCs presence in the debris. Some classes of compounds (furans, sulphur-containing species) showed weak interactions with the tested materials and were relatively quickly removed from the surroundings of the urine samples. Conversely, more polar analytes (e.g., ketones) were more influenced [14]. Predictably, the increase of molecular mass promoted the interactions with debris and increased the residence times of VOCs in void spaces [14]. Huo *et. al.* [15] monitored species released by healthy volunteers closed in an environmental chamber mimicking void space and permeating through a glass column packed with different discs of building materials. The study involving the whole body emission demonstrated the permeation of $CO_2$, $NH_3$, acetone, and isoprene through a collapsed building simulator.

Urban air is typically highly contaminated with numerous volatile organic compounds. They predominantly stem from anthropogenic sources such as vehicle exhausts, solvents and fuel evaporation, fossil fuel combustion, or emissions of liquefied petroleum gas [90-93]. Their levels may vary over relatively brief periods of time, show diurnal/seasonal cycles, or exhibit spikes related to local temporal emissions. Moreover, the profiles of urban VOCs' differ from one country to another due to differences in heating patterns, composition of vehicle fuel, local regulations concerning VOCs emissions, or climatic conditions. This highly complex and variable phase becomes even more complicated and harsh after massive buildings collapse. Damaged building structures, sewage systems, broken gas pipes, fire and smoke produce additional contaminants and/or toxic agents, which mix with air filling the void spaces and human-borne volatiles [17, 18]. In particular, released toxic agents might embrace polychlorinated biphenyls, hazardous metals, asbestos, various harmful gases (e.g., hydrogen cyanide, hydrogen sulfide, halogenated gases, carbon monoxide), detergents, acids and alkalis, ethylene glycol, propylene glycol, phenol, and alcohols [94, 95]. Furthermore, volatiles emitted by rodents/insects, or decomposing bodies can additionally complicate the chemical analysis at the disaster site and induce false readings [96]. All these factors and confounders considerably affect the levels of the human-specific volatiles in the voids of collapsed buildings and make their identification and detection a really challenging task.

The complexity and unpredictability of an entrapment environment, variety of confounders and interactions as well as ethical restrictions limit also the laboratory-based studies in this specific field. Thus, it is very difficult to model in reliable way the behavior of the human-specific chemical fingerprint in the surroundings of the entrapped person and predict the levels of its constituents. On the other hand, the knowledge of even approximate concentrations of potential indicators of human presence would provide invaluable profits for the chemical analysis towards victim location. These include (i) validation of potential markers against the possibility of their detection in highly polluted air in the disaster environment, (ii) selection of appropriate analytical instruments, which could be used for the field detection of entombed victims and (iii) optimization of these techniques for the detection of human markers in the disaster environment.

As mentioned above, air filling void spaces in the collapsed buildings is highly contaminated with a complex chemical signature. In particular it is characterized by variable and unpredictable levels of VOCs, which can interfere with human-specific chemical fingerprints. Thus, debris air constitutes a background, on which markers of human presence have to be identified and detected. Despite these limitations an effort was made within this work to tentatively verify the preselected markers against their typical urban levels. Such a comparison could help, for example, to exclude markers exhibiting emission rates, which are too small to provide air levels, which can be reliably distinguished from the background. For this purpose a simple model of the dispersion of human-borne VOCs in debris has been developed. In this model an entrapped human is represented by a ball (i.e., a radially symmetric structure with a certain radius $R$) emitting a constant stream $j$ of VOCs. Volatiles are assumed to be inert species, which do not interact with the debris in the disaster environment (no losses/sinks of VOCs). Moreover, the transport of VOCs in the voids of collapsed building is restricted only to diffusion and the diffusion coefficients are postulated to be homogeneous and constant. The results of the diffusion calculation are independent from the specifically chosen radius of this ball (assuming, of course, that the distance from the center of the ball is larger than the radius $R$). A detailed description of the applied model and the calculations of tentative levels of human markers in the vicinity of an entrapped victim are given in Appendix A. In general the model predicts that VOC concentration decrease proportional to the inverse of the distance from the victim. It should be stressed here that although such a model is unrealistic and provides presumably overestimated concentrations of species of interest it can be used as the first verification tool for the proposed preliminary markers. The exemplary calculations done for the arbitrary chosen conditions: distance of 3 m from the entrapped person and a debris-to-air ratio 3:1 are presented in Table 2 and illustrated in Figure 4. Hence we assume that the volume of debris is three times the volume of air, which is a reasonable assumption (and also can easily be adapted, see formula (A.17) in the Appendix A). The realistic estimation of the latter factor poses an additional challenge, as it depends on the collapse pattern, which in term is determined by the type of the building construction and the building code [5, 6]. Due to the shortage of information on this parameter it is futile to indicate its typical or most probable value. However, it is reasonable to assume that it will be smaller for well-constructed, reinforced concrete buildings. The values presented in Table 2 should be treated rather as upper boundaries of possible marker levels. For these particular conditions the majority of compounds is expected to exhibit low-ppb levels. More specifically, concentrations of 19 species (40%) might be spread around 1 ppb and levels of further 13 (28%) should not exceed 10 ppb. Only seven compounds ($CO_2$, CO, ammonia, acetone, methanol, ethanol, and acetaldehyde) could produce levels higher than 100 ppb in the vicinity of survivors. These values can be next compared to the typical urban concentrations of species of interest. For this purpose an extensive literature search was done. An effort was made to select data reported for cities from different countries and continents to include region/continent dependent differences in concentrations. It should be stressed that in the case of several compounds the urban/indoor air data are difficult to obtain. This is due to their low toxicity, ultra-small levels and, consequently, lack of regulations concerning their emissions. The typical urban air levels of markers under scrutiny have been listed in Table 2 and are shown in Figure 4. In general, we believe that several valuable pieces of information can be extracted from the juxtaposition of the aforementioned data. First, levels of several tentative indicators can be too low to be distinguished from the background in the void spaces of collapsed buildings. For instance, emission of CO from the human body can produce levels which are comparable to the ones usually occurring in urban air. Bearing in mind that CO levels exhibit diurnal and seasonal variability it can be very difficult to separate urban and human-borne components of CO levels during a field chemical analysis. Moreover, concentrations of CO can change considerably during the USaR operation as a result of incidental hazards or fires [18, 19]. The same holds true for methanol and ethanol. Although both of these alcohols show relatively high abundances in the human chemical signature, their urban levels are also high. The high urban background stems from the increasing use of alcohols as alternative energy sources replacing gasoline and related emission of unburned fuel, or their evaporation from leaking tanks [97]. Overall, the suitability of several constituents from the proposed set can be reduced by high and variable urban air levels. Bearing in mind all aforementioned problems and confounders it is difficult to establish a clear criterion, which would exclude markers being too affected by urban air to be applied for a detection of entrapped victims. Nevertheless, a sufficiently high difference between background levels and human-borne levels in debris air seems to be a reasonable discriminant. Within this review, voids concentrations at least ten times higher than urban air background were recognized as a threshold. Several species from the original set failed to fulfill this criterion. These were CO, ethanol, 2-propanol, NO, DL-limonene, n-nonane, n-octane, α-pinene, and n-heptane. Interestingly, apart from DL-limonene and α-pinene species from this group are typical vehicle exhausts, or fuel vapors. This finding renders the vehicle-related pollution one of the main confounders hindering the application of the human chemical fingerprint during USaR operations.

## 6. Analytical instrumentation for field VOCs detection

The lab-based analytical instruments commonly used to determine and track volatile species forming the human-specific chemical pattern are inherently large in size, expensive, demand laborious and time-consuming sample-preparation methods, and require well-trained and experienced operators. This places significant limitations on their routine use in field conditions in general and disaster environment in particular. Here, simple-in-use ("yes/no" response), rapid, hand-held, low-energy and simultaneously sensitive screening instruments are desirable. A number of technologies could meet these requirements. Recent rapid progress in electronic sensor technology has stimulated the development of devices known as electronic noses (e-noses) [98]. These instruments are arrays of different non-selective sensors capable of detecting and discriminating a wide diversity of chemical species. Strictly speaking, their responses are not correlated to one specific compound, but rather to the whole chemical fingerprint. Thus, e-noses discriminate different VOCs' profiles using qualitative or semi-quantitative information. The versatile capabilities of e-noses stem from the variety of sensors available for selection for sensor arrays (i.e., conducting-polymer sensors, metal oxide sensors, optical sensors, electrochemical sensors) and abilities of manufacturers to produce customized, low-cost, and multi-use devices for particular applications [98-101]. However, a key prerequisite for the success of e-nose devices in a particular application is the knowledge of the chemical pattern of interest, which can only be provided by more sophisticated analytical techniques. If successful, sensor arrays may revolutionize USaR operations, when built into robust and small instruments. Once installed, e.g. on borescopes they could penetrate the collapsed structures and screen their interiors for volatile chemical signs-of-life. Despite encouraging facilities sensor arrays suffer from several disadvantages such as temperature-dependent stability, or humidity effects. Another promising technique is ion mobility spectrometry (IMS) separating volatiles on the basis of differences in their migration speed in an inert buffer gas under the influence of an electric field [102]. Recent rapid advances in ion mobility spectrometry resulted in the development of numerous sub-techniques exploiting different strengths and forms of the electric fields (e.g., linear drift tube IMS (DTIMS), travelling wave IMS (TWIMS), aspiration IMS (AIMS), or field-asymmetric IMS (FAIMS)), or combining IMS with other techniques such as, e.g. gas chromatography (multi-capillary column IMS (MCC-IMS)). The IMS instruments can be miniaturized, measure rapidly, have low energy consumption, and are very sensitive. Moreover, these techniques have already been successfully applied for the field detection and identification of chemical warfare agents (CWA), or toxic industrial chemicals (TICs), and the expertise and know-how gained within these applications could be transferred into the "search and rescue" science notably accelerating the pace of investigations. Fast gas chromatography (fast GC) combined with mass spectrometric detection also exhibits a considerable potential for USaR operations. Short analysis time (several minutes) and progressive miniaturization render this technique a perspective locating tool of entrapped victims. The fast GC-MS instruments can be field-portable [103, 104] and in combination with some pre-concentration methods (e.g., solid phase microextraction SPME) could provide the detection of the majority of volatiles under scrutiny. Nevertheless, their weight (10-20 kg) still hinders their applicability during searching large disaster areas.

The successful employment of the aforementioned techniques for the location of entrapped victims is, however, determined by their analytical limitations. Here the limit of detection (LOD) can be regarded as a basic factor influencing the selection of the optimal technique. Table 2 lists exemplary LODs of different sensor-, or IMS-based instruments reported for some of compounds of interest. It should be stressed here, that some of these analytical tools are still at the early phase of the development and should be considered as prototypes. Several conclusions can be distilled from this comparison. First, the detection of many potential volatile signs-of-life can pose a challenge due to their ultra-low concentrations and unsatisfactory capabilities of the available field techniques. Nevertheless, the rapid progress in analytical chemistry instrumentation is expected to solve this problem in the future. In this context an interesting technique is multi-capillary column ion mobility spectrometry. Although this technique is strictly speaking not hand-held and real-time, and can impose operational problems for unexperienced users, it offers LODs, which are adequate for the detection of many compounds under scrutiny. Indeed, recently MCC-IMS was demonstrated to be capable of detecting some constituents of the human scent [13].

## 7. Classification of potential markers of human presence

An ideal marker of human presence should be omnipresent, volatile, relatively non-reactive, continuously emitted by the human body and present at relatively high concentrations in the proximity of an entrapped victim. In this spirit, we propose the classification of potential indicators of human presence preselected within this review into three subsets. Subset A comprises predominantly endogenous species exhibiting high emission rates from human body, which can produce debris concentrations detectable by currently available, portable field analyzers and are clearly distinguishable from urban background levels. This group represents the most promising markers. Subset B contains volatiles of different (frequently unknown) origins with tentatively predicted debris levels being at least ten-fold higher than the expected background levels, however, too low to be reliably detected by current portable techniques. The category C incorporates volatiles stemming from sources, which can be suppressed during the entrapment, or species with potential debris levels difficult to separate from urban background. The proposed classification of particular species of interest has been applied in Table 2. Following this classification $CO_2$, ammonia, acetone, 6-methyl-5-hepten-2-one, isoprene, n-propanal, n-hexanal, n-heptanal, n-octanal, n-nonanal, and acetaldehyde constitute the class A and thereby are the strongest candidates for markers of human presence. The potential of the class A species is additionally supported by the fact, that some of them have already been indicated as promising human indicators by several early studies [13-15].

## 8. Conclusions

One of the main goals of this review was to create a database of human-borne volatiles having a high potential as markers of human presence, which could be used for early location of entrapped victims during rescue operations and to estimate their emission rates from the human body on the basis of existing literature data. Altogether 47 compounds were pre-selected using skin emission and quantitative exhaled breath data. They belong to several chemical classes; however, aldehydes and hydrocarbons are the most numerous ones. It should be stressed that these species may originate from several distinct sources and their production is still far from being completely understood. Due to the nature of this specific field, ethical and methodological restrictions the prediction of fluxes of these species and their concentrations in the voids of collapsed buildings poses considerable challenges and problems. In particular, unpredictable and variable conditions in the entrapment environment, shortage of quantitative data on the VOCs' emission by the human body, or poorly known interactions of VOCs with debris materials affect efforts towards this goal. In this context, the emission rates calculated within this work should be treated as tentative and the predicted concentrations in void spaces as approximate indicating only the order of magnitude of the expected levels appearing in real situations (e.g., low ppb, ppt). We believe that several valuable pieces of information can be distilled from the data presented in this work. First, optimal field analytical techniques can be selected on the basis of the markers' physicochemical characteristics and their approximate levels in the confined spaces. These techniques can be further improved to provide an optimal response at the disaster site (targeted analysis). Moreover, species from the proposed set can be verified by lowering the value of those, which produce a signal too small to be reliably separated from the background.

To sum up, the set of potential markers of the human presence preselected within this review should be considered as an initial database of species to be verified during field studies. Within this context, a major focus lies on the investigations of interactions of volatiles of interest with building materials and other adsorbents in the disaster environment such as clothing, dust, or soil. Further efforts will need to take into account disaster-related emission of species as well as different conditions such as, e.g. temperature and humidity affecting the surface chemistry. Thus, we expect that the VOC database proposed here will be further complemented and verified. The success of chemical analysis toward the detection of humans will, however, primarily depend on the availability of analytical technologies for the rapid, continuous, and field detection of volatiles as signs-of-life. Although a number of techniques show a huge potential in this context, their applicability has to be verified in harsh, highly contaminated, and toxic disaster environment. In this sense, we recognize that data and considerations included in this review will guide future investigations in this exciting field.

**Acknowledgments**

We appreciate funding from the Austrian Federal Ministry for Transport, Innovation, and Technology (BMVIT/BMWA, project 836308, KIRAS). P.M. and K.U. gratefully acknowledge support from the Austrian Science Fund (FWF) under Grant No. P24736-B23. G.T. also gratefully acknowledges support from the Austrian Science Fund (FWF) under Grant No. Y330. We thank the government of Vorarlberg (Austria) for its generous support.

**Appendix A. On decrease of exhaled VOC concentrations**

We will use the following notation: $t$ time variable, $x$ space coordinates, $C(t,x)$ concentration of a VOC, $V$ volume, $S$ surface of $V$, $\partial V$ boundary of $V$, $F$ flux, $n$ normal vector.

For the following compare Evans, Chapter 2.3 (L. C. Evans, *Partial Differential Equations*, 2nd ed., Amer. Math. Soc., Providence, 2010)

The change of mass in a volume $V$ is given by the flux through the surface of $V$ plus the net production in $V$. Thus the mass balance for a fixed volume $V$ then reads

$$\frac{d}{dt}\int_V C(t,x)\,dV = -\int_{\partial V} F\cdot n\,dS + \int_V f(t,x)\,dV \tag{A.1}$$

where $f$ is the production rate in $V$. Using Stokes' theorem (Gauss's divergence theorem) which states

$$\int_{\partial V} F\cdot n\,dS = \int_V (\nabla\cdot F)\,dV \tag{A.2}$$

(where $\nabla .\ F = \text{div}\,F$) we arrive at

$$\frac{d}{dt}\int_V C(t,x)\,dV = -\int_V (\nabla\cdot F)\,dV + \int_V f(t,x)\,dV. \tag{A.3}$$

By Fick's law the flux $F$ is proportional (diffusion constant $a > 0$) to the gradient of the concentration $\nabla\, C(t,x)$.

$$F = -a\nabla C(t,x). \tag{A.4}$$

We arrive at

$$C_t(t,x) = \nabla\cdot(a\nabla C(t,x)) + f(t,x). \tag{A.5}$$

Remark: up to this point the derivation is completely general. We make now the following assumptions:

Assumption 1: $a$ is homogeneous and constant

Assumption 2: A human is modeled by a ball with radius $R$ emitting a constant stream of a VOC (e.g., nmol/min) of the form

$$f(t,x) = f(x) = \frac{3\dot{J}}{4\pi R^3}\chi_{B_R(0)}(x), \tag{A.6}$$

where $\chi_{B_{R(0)}}$ denotes the characteristic function of a ball with radius $R$ at $x = 0$ and $\dot{J}$ the emitted stream. Here $f$ is chosen such that there is constant production within the ball which totals to $\dot{J}$. Note that the specific form of $f$ will not affect the concentration outside the ball as long as it is radially symmetric.

Then we have

$$C_t(t,x) = a\Delta C(t,x) + f(x). \tag{A.7}$$

Now we consider stationary solutions

$$0 = a\Delta C(t,x) + f(x). \tag{A.8}$$

A special solution $C_s$ of the non-homogeneous Equation (A.8) is given by

$$C_s(x) = \frac{9\dot{J}}{4\pi Ra}\begin{cases} \frac{1}{2}(1-\frac{|x|^2}{3R^2}), & |x| \leq R, \\ \frac{1}{3}\frac{R}{|x|}, & |x| \geq R. \end{cases} \tag{A.9}$$

If we consider an initial concentration

$$C(0,x) = g(x) \tag{A.10}$$

the general solution of Equation (A.7) is then given by

$$C(t,x) = C_s(x) + (\Phi_t * \tilde{g})(x), \quad \tilde{g}(x) = g(x) - C_s(x), \tag{A.11}$$

where

$$\Phi_t(x) = \frac{1}{(4\pi\sqrt{at})^{3/2}}\mathrm{e}^{-\frac{|x|^2}{4\sqrt{at}}} \tag{A.12}$$

denotes the fundamental solution of the heat equation and $*$ denotes convolution.

Note that the convergence to the equilibrium concentration can be estimated by

$$||\Phi_t * \tilde{g}||_\infty \leq \frac{1}{(4\pi\sqrt{at})^{3/2}}||\tilde{g}||_\infty \tag{A.13}$$

and in the special case $g \equiv 0$, that is $\tilde{g} = -C_s$ , we have

$$||\tilde{g}||_\infty = ||C_s||_\infty = \frac{3\dot{J}}{4\pi aR}. \tag{A.14}$$

**Summary**: the stationary solution reads

$$C_s(x) = \frac{3\dot{J}}{4\pi a}\frac{1}{|x|}, \qquad |x| \geq R. \tag{A.15}$$

An example: $\dot{J} = 12$ nmol/min, $a \approx 0.1$ cm$^2$/sec = 0.0006 m$^2$/min yields

$$C_s(x) = 4775\ [nmol/m^2]\ \frac{1}{|x|}, \qquad |x| \geq R. \tag{A.16}$$

At a distance of 10 m this yields 477.5 nmol/m$^3$ or 0.4775 nmol/l.

Remark 1: Inert debris will raise the concentration and can be taken into account at the first attempt by scaling the distance $|x|$ by a factor $\rho^{1/3}$, $0 < \rho = (V_0 - V_f)/V_0 \leq 1$, where $V_0$ is the volume in the absence of debris and $V_f$ is the volume filled by debris.

$$C_s(x) = \frac{3\dot{J}}{4\pi a}\frac{1}{\rho^{1/3}}\frac{1}{|x|}, \qquad |x| \geq R. \tag{A.17}$$

If we assume that the volume of debris is three times the volume of air then

$V_0 := V_{debris} + V_{air}$ , $V_f := V_{debris} = 3V_{air}$. This yields $\rho = 1/4$ and $1/\rho^{1/3} \approx 1.59$.

Remark 2: In our simple model we assumed that diffusion is constant and homogenous which allowed for an analytical solution of the problem. Other geometries can be incorporated and computed numerically.

Remark 3: Initially when a human is suddenly entrapped the surrounding will show the background level of VOCs. As he/she emits VOCs these VOCs will diffuse into his/her surrounding and after a certain time a constant distribution of VOCs will be established according to formula (A.17). The concentration of these VOCs is proportional to the stream $\dot{J}$ he emits and will decrease proportional to the inverse of the distance $|x|$ from him/her and is also inverse proportional to $a$ where $a$ is the diffusion constant.

If there are two or three persons close together within a ball of radius $R$ then one can simply multiply the stream $\dot{J}$ by a factor of 2 or 3.

Table 1. Breath concentrations, skin emissions, tentative origins and whole body fluxes of potential volatile markers of human presence. Urine and blood omnipresent species taken from [16, 28, 43, 46].

| A | B | C | D | E | F | G | H | I |
|---|---|---|---|---|---|---|---|---|
| Compound<br>CAS | Breath levels<br>mean (population)<br>[ppb] | Skin emission<br>mean (population) | Urine | Blood | Tentative Origin in humans | Flux$_{breath}$<br>[nmol/min] | Flux$_{skin}$<br>[nmol/min] | Flux$_{total}$<br>[nmol/min] |
| $CO_2$<br>124-38-9 | (a) 4.9 % (19) [105]<br>(b) 6.1 % (6) [44] | (a) $3.4\times10^{-5}$ ml×cm$^{-2}$×min$^{-1}$ (63) arm/hand [31] | | ● | (a) cellular respiration | $66.8\times10^{5}$ | $26.5\times10^{3}$ | $67.1\times10^{5}$ |
| NO<br>10102-43-9 | (c) 7.8 (294)[106]<br>(d) 7.2 (20)[107]<br>(e) 8.2 (10)[108]<br>(f) 27.6 (106)[109]<br>(g) 18.9 (26)[110]<br>(h) 17.5 (89)[111] | (b) 12.8 fmol×cm$^{-2}$×min$^{-1}$ (14) hand[30]<br>(c) 79.5 fmol×cm$^{-2}$×min$^{-1}$ (14) hand/arm[35] | ● | | (a) Enzymatic oxidation of L-arginine (iNOS)[112] | 1.8 | 0.8 | 2.6 |
| CO<br>630-08-0 | (a) 3.2 ppm (20)[107]<br>(b) 2.9 ppm (37)[113]<br>(c) 4.3 ppm (239)[114]<br>(d) 3.6 ppm (55)[115]<br>(e) 4.1 ppm (857)[116] | | | ● | (a) Hemoprotein turnover[45] | 524.5 | | 524.5 |
| Ammonia<br>7664-41-7 | (a) 1015 (5)[117]<br>(b) 854 (17)[118]<br>(c) 589 (48)[119]<br>(d) 775 (20)[33]<br>(e) 480 (30)[48] | (a) 0.5 ng×cm$^{-2}$×min$^{-1}$ (30) hand[33] | ● | ● | (a) Bacterial metabolism of proteins in gut [42]<br>(b) Bacterial metabolism of proteins in oral cavity [43, 120] | 90.9 | 513.8 | 604.7 |
| Acetone<br>67-64-1 | (a) 487 (5)[117]<br>(b) 477 (30)[48]<br>(c) 456 (17)[118]<br>(d) 226 (143)[121]<br>(e) 255 (31)[122]<br>(f) 217 (40)[123]<br>(g) 950 (28)[46]<br>(h) 628 (215)[47] | (a) 1370 fmol×cm$^{-2}$×min$^{-1}$ (31) hand [29]<br>(b) 44.8 nmol×person$^{-1}$×min$^{-1}$ (10) body[32]<br>(c) 4.3 ng×cm$^{-2}$×h$^{-1}$ (60) hand[34] | ● | ● | (a) endogenous decarboxylation of Acetyl–CoA[47]<br>(b) oxidation of squalene [51]<br>(c) 2-propanol metabolism [52]<br>(d) diet | 59.8 | 25 | 84.8 |
| 2-Butanone<br>78-93-3 | (a) 5.1 (143)[121]<br>(b) 0.24 (40)[123]<br>(c) 2.6 (28)[46] | (a) 7.2 fmol×cm$^{-2}$×min$^{-1}$ (31) hand[29]<br>(b) 4.3 nmol×person$^{-1}$×min$^{-1}$ (10) body[32] | ● | ● | (a) diet[68] | 0.5 | 1.15 | 1.65 |
| 2-Pentanone<br>107-87-9 | (a) 4.8 (143)[121]<br>(b) 0.36 (40)[123]<br>(c) 0.62 (28)[46]<br>(d) 0.22 (7)[124] | (a) 2.47 fmol×cm$^{-2}$×min$^{-1}$ (31) hand[29] | ● | ● | (a) diet[68]<br>(b) 2-pentanol metabolism[125] | 0.43 | 0.05 | 0.47 |
| 5-Hepten-2-one, 6-methyl-<br>110-93-0 | | (a) 212.9 fmol×cm$^{-2}$×min$^{-1}$ (31)hand[29]<br>(b) 0.98 nmol×person$^{-1}$×min$^{-1}$ (10) body[32] | | | (a) cutaneous oxidation of squalene [51] | | 3.08 | 3.08 |
| 3-Buten-2-one<br>78-94-4 | (a) 3.8 (28)[46]<br>(b) 5.5 (143)[121] | (a) 9.2 fmol×cm$^{-2}$×min$^{-1}$ (31) hand[29]<br>(b) 6.8 nmol×person$^{-1}$×min$^{-1}$ (10) body[32] | | | (a) oxidation of isoprene[83] | 0.67 | 1.78 | 2.45 |
| 2,3 Butanedione<br>431-03-8 | (a) 29 (28)[46] | | | | (a) diet (butter)[68] | 3.74 | | 3.74 |
| Acetaldehyde<br>75-07-0 | (a) 67.4 (143)[121]<br>(b) 5.5 (12)[126]<br>(c) 24 (30)[127] | (a) 466 fmol×cm$^{-2}$×min$^{-1}$ (31) hand [29]<br>(b) 3.8 ng×cm$^{-2}$×h$^{-1}$ (60) hand [34] | ● | ● | (a) ethanol metabolism[63]<br>(b) cutaneous oxidative degradation of linolenic acid[59] | 7.3 | 15.4 | 22.7 |
| n-Propanal<br>123-38-6 | (a) 18.3 (28)[46]<br>(b) 6.9 (143)[121] | (a) 18.4 fmol×cm$^{-2}$×min$^{-1}$ (31)hand[29]<br>(b) 6.6 nmol×person$^{-1}$×min$^{-1}$ (10) body[32] | ● | | (a) cutaneous of linolenic acid and oleic acid[59]<br>(b) 1-propanol metabolism<br>(c) diet[68] | 1.13 | 1.85 | 2.98 |
| 2-Propenal<br>107-02-8 | (a) 5.9 (28)[46] | (a) 21 fmol×cm$^{-2}$×min$^{-1}$ (31) hand[29] | | | (a) smoking[85] | 0.76 | 0.37 | 1.13 |
| 2-Propenal, 2-methyl<br>78-85-3 | (a) 1.2 (28)[46] | (a) 20.5 fmol×cm$^{-2}$×min$^{-1}$ (31)hand[29]<br>(b) 0.54 nmol×person$^{-1}$×min$^{-1}$ (10) body[32] | | | (a) oxidation of isoprene[83] | 0.15 | 0.42 | 0.57 |
| Propanal, 2-methyl-<br>78-84-2 | | (a) 11.6 fmol×cm$^{-2}$×min$^{-1}$ (31)hand[29] | ● | | (a) diet [128] | | 0.21 | 0.21 |
| Butanal, 2-methyl-<br>96-17-3 | | (a) 13.9 fmol×cm$^{-2}$×min$^{-1}$ (31) hand[29] | ● | | (a) diet [128] | | 0.25 | 0.25 |
| Butanal, 3-methyl-<br>590-86-3 | | (a) 15.1 fmol×cm$^{-2}$×min$^{-1}$ (31)hand[29] | ● | | (a) diet [128] | | 0.28 | 0.28 |
| n-Hexanal<br>66-25-1 | (a) 15.4 (31)[122] | (a) 56.2 fmol×cm$^{-2}$×min$^{-1}$ (31) hand[29]<br>(b) 2.46 nmol×person$^{-1}$×min$^{-1}$ (10) body[32] | ● | | (a) cutaneous degradation of linoleic, palmitoleic and vaccenic acids[60] | 2.1 | 1.36 | 3.44 |
| n-Heptanal<br>111-71-7 | (a) 0.07 (12)[126] | (a) 29.8 fmol×cm$^{-2}$×min$^{-1}$ (31) hand[29]<br>(b) 1.85 nmol×person$^{-1}$×min$^{-1}$ (10) body[32] | ● | | (a) Cutaneous oxidative degradation of palmitoleic acid, vaccenic acid[60] | <0.001 | 0.84 | 0.84 |
| n-Octanal<br>124-13-0 | (a) 0.27 (12)[126] | (a) 42.7 fmol×cm$^{-2}$×min$^{-1}$ (31) hand[29]<br>(b) 1.3 nmol×person$^{-1}$×min$^{-1}$ (10) body[32] | ● | | (a) oxidative degradation of oleic acid[59] | 0.04 | 0.88 | 0.92 |
| n-Nonanal<br>124-19-6 | (a) 0.8 [126] | (a) 60.2 fmol×cm$^{-2}$×min$^{-1}$ (31) hand[29]<br>(b) 2.16 nmol×person$^{-1}$×min$^{-1}$ (10)[32] | ● | | (a) oxidative degradation of oleic acid[60] | 0.11 | 1.33 | 1.44 |
| Isoprene<br>78-79-5 | (a) 89 (5)[117]<br>(b) 118 (30)[129]<br>(c) 99.3 (205)[64]<br>(d) 71 (143)[121]<br>(e) 131 (28)[46] | (a) 4.6 fmol×cm$^{-2}$×min$^{-1}$ (31) hand [29] | ● | ● | (a) endogenous cholesterol synthesis[64]<br>(b) peroxidation of squalene[55]<br>(c) cutaneous synthesis of squalene[130] | 12.0 | 0.09 | 12.1 |
| 1,3-Pentadiene, 2- | | (a) 2.54 fmol×cm$^{-2}$×min$^{-1}$ (31) | | | | | 0.05 | 0.05 |
| methyl-, Z-<br>2787-45-3 | | hand[29] | | | | | | |
| 1,3-Pentadiene, 2-methyl-, E-<br>926-54-5 | | (a) 1.7 fmol$\times$cm$^{-2}\times$min$^{-1}$ (31) hand[29] | | | | | 0.03 | 0.03 |
| 2-Pentene, 2-methyl-<br>625-27-4 | | (a) 12.7 fmol$\times$cm$^{-2}\times$min$^{-1}$ (31)hand[29]<br>(b) 0.32 nmol$\times$person$^{-1}\times$min$^{-1}$ (10) body[32] | | | (a) peroxidation of squalene[55] | | 0.25 | 0.25 |
| 1-Heptene<br>592-76-7 | | (a) 1.73 fmol$\times$cm$^{-2}\times$min$^{-1}$ (31)hand[29] | | | | | 0.003 | 0.003 |
| n-Heptane<br>142-82-5 | | (a) 3.3 fmol$\times$cm$^{-2}\times$min$^{-1}$ (31)hand[29] | | | (a) cutaneous degradation of oleic acid[59] | | 0.06 | 0.06 |
| 1-Octene<br>111-66-0 | | (a) 3.25 fmol$\times$cm$^{-2}\times$min$^{-1}$ (31)hand[29] | | | | | 0.06 | 0.06 |
| n-Octane<br>111-65-9 | (a) 0.12 (28)[46] | (a) 8.3 fmol$\times$cm$^{-2}\times$min$^{-1}$ (31) hand[29] | | ● | (a) oxidative degradation of oleic acid[59] | 0.02 | 0.15 | 0.17 |
| 1-Nonene<br>124-11-8 | | (a) 3.7 fmol$\times$cm$^{-2}\times$min$^{-1}$ (31)hand[29] | | | | | 0.06 | 0.06 |
| n-Nonane<br>111-84-2 | | (a) 14.3 fmol$\times$cm$^{-2}\times$min$^{-1}$ (31)hand[29] | | | | | 0.26 | 0.26 |
| Methanol<br>67-56-1 | (a) 450 (30)[131]<br>(b) 272 (20)[132]<br>(c) 202 (10)[133] | (a) Emitted (no quantitative data)[67, 134] | ● | ● | (a) bacterial metabolism of carbohydrates in gut [69]<br>(b) diet | 45.1 | * | 45.1 |
| Ethanol<br>64-17-5 | (a) 86 (5)[117]<br>(b) 189 (143)[121]<br>(c) 196 (30)[127]<br>(d) 233 (15)[135]<br>(e) 165 (20)[132]<br>(f) 46 (15)[136] | (a) Emitted (no quantitative data)[67] | ● | ● | (a) gut bacterial metabolism[41]<br>(b) diet | 23.1 | * | 23.1 |
| 2-Propanol<br>67-63-0 | (a) 22 (30)[48]<br>(b) 150 (46)[137] | (a) Emitted (no quantitative data)[67] | | | (a) diet[68]<br>(b) acetone metabolism[73] | 12.84 | * | 12.84 |
| Dimethylsulfide<br>75-18-3 | (a) 9.3 (143)[121]<br>(b) 35 (50)[138]<br>(c) 7.3 (31)[122]<br>(d) 13.9 (40)[123]<br>(e) 5 (28)[46]<br>(f) 7.6 (20)[132] | (a) 2.52 fmol$\times$cm$^{-2}\times$min$^{-1}$ (31) hand[29] | ● | ● | (a) endogenous metabolism of sulfur-containing amino acids[76]<br>(b) bacterial decomposition of sulfur-containing amino acids[76] | 1.77 | 0.05 | 1.81 |
| Allyl methyl sulfide<br>10152-76-8 | (a) 0.1 (40)[123]<br>(b) 1.6 (28)[46] | | | | (a) Diet (garlic)[74] | 0.09 | | 0.09 |
| Methyl propyl sulfide<br>3877-15-4 | (a) 2.6 (28)[46] | | | ● | (a) Diet (onion)[75] | 0.29 | | 0.29 |
| p-Cymene<br>99-87-6 | (a) 0.14 (28)[46] | (a) 4.9 fmol$\times$cm$^{-2}\times$min$^{-1}$ (31) hand[29] | | ● | (a) diet[68] | 0.018 | 0.075 | 0.094 |
| DL-Limonene<br>138-86-3 | (a) 1.46 (28)[46]<br>(b) 2.3 (20)[132] | (a) 25.6 fmol$\times$cm$^{-2}\times$min$^{-1}$ (31) hand[29] | ● | ● | (a) diet[68] | 0.23 | 0.54 | 0.77 |
| | | (b) 0.89 nmol×person$^{-1}$×min$^{-1}$ (10) body[32] | | | | | | |
| α-Pinene<br>80-56-8 | (a) 0.6 (28)[46] | | | | (a) perfumes, cosmetics | 0.08 | | 0.08 |
| Furan, 2-methyl-<br>534-22-5 | (a) 0.55 (28)[46]<br>(b) 9.5 (143)[121] | (a) 1.9 fmol×cm$^{-2}$×min$^{-1}$ (31) hand[29] | ● | | (a) smoking[85] | 1.03 | 0.03 | 1.06 |
| Furan, 3-methyl-<br>930-27-8 | (a) 0.18 (28)[46] | (a) 1.2 fmol×cm$^{-2}$×min$^{-1}$ (31) hand[29] | ● | ● | (a) oxidation of isoprene[83]<br>(b) skin microbiota metabolism[81] | 0.023 | 0.023 | 0.045 |
| Furan, 2-pentyl-<br>3777-69-3 | | (a) 2.3 fmol×cm$^{-2}$×min$^{-1}$ (31)hand[29] | ● | | (a) Cutaneous oxidation of linolenic acid[139]<br>(b) skin microbiota metabolism[80] | | 0.04 | 0.04 |
| Acetonitrile<br>75-05-8 | (a) 31.5 (28)[46]<br>(b) 2.0 (19)[86]<br>(c) 5.7 (77)[87] | (a) 26.8 fmol×cm$^{-2}$×min$^{-1}$ (31) hand[29] | ● | ● | (a) smoking[85] | 1.41 | 0.56 | 1.98 |
| γ-Butyrolactone<br>96-48-0 | (a) 2.8 (28)[46] | (a) 34.4 fmol×cm$^{-2}$×min$^{-1}$ (31)hand[29] | | | (a) skin microbiota metabolism[79]<br>(b) diet[68] | 0.36 | 0.59 | 0.95 |
| Methyl acetate<br>79-20-9 | (a) 2.6 (28)[46]<br>(b) 0.98 (7)[124] | | ● | ● | | 0.29 | | 0.29 |
| Dimethyl selenide<br>593-79-3 | (a) 0.35 (28)[46]<br>(b) 0.13 (40)[123] | | | ● | (a) selenomethionine and selenocysteine metabolism | 0.029 | | 0.029 |

Table 2 Predicted levels in void spaces (see text for detailed conditions), exemplary urban levels, possibilities of detection by portable instruments and classification of potential volatile markers of human presence. The classification criteria are defined in section 6.

| A | B | C | D | E | F |
|---|---|---|---|---|---|
| Compound<br>CAS | Predicted level at 3 m distance [ppb] | Exemplary urban air levels | Main urban sources | Detection possibilities<br>LOD, technique | Class of marker |
| $CO_2$<br>124-38-9 | 30×10$^{6}$ | (b) 408 ppm (Dallas) [140]<br>(c) 390 ppm (Phenix)[141]<br>(d) 469 ppm (Wrocław)[142]<br>(e) 403-408 ppm (Portland)[143] | Vehicle emissions, | (a) 0.25% optical sensor [92]<br>(b) 0.23% Solvatochromic probe [144] | A |
| NO<br>10102-43-9 | 7.7 | (a) 24.5 ppb (Hong Kong)[145]<br>(b) 11.7 ppb (A Coruna)[146]<br>(c) 127/35 ppb (Seul)[147] | vehicle exhaust | (a) 6 ppb chemiresistor (PEDOT:PSS/$TiO_2$)[148]<br>(b) 5 ppb electrochemical ($WO_3$/Pt)[149]<br>(c) 18 ppb chemiresistor ($WO_3$/$Cr_2O_3$)[150]<br>(d) 3.6 ppb ICOS[151]<br>(e) 0.03 ppb lase QCL[152]<br>(f) 4 ppb electrochemical sensor[153] | C |
| CO<br>630-08-0 | 1350 | (a) 1.6 ppm (Karachi)[90]<br>(b) 0.592 ppm (Hong Kong)[145]<br>(c) 1.7 ppm (Rio de Janeiro)[154]<br>(d) 0.53 ppm (London) [155] | Coal burning<br>Vehicular exhost<br>Cigarette smoke | (a) 1 ppm chemiresistor (Ca-$SnO_2$)[156]<br>(b) 1 ppm electrochemical sensor [157]<br>(c) 0.1 ppm controlled potential electrolysis[158]<br>(d) 4 ppb electrochemical sensor[153] | C |
| | | (e) 1.2 ppm (Seul)[147] | | | |
| Ammonia<br>7664-41-7 | 1260 | (a) 22 ppb (Santiago, Chile)[159]<br>(b) 24.7 ppb (Rome, Italy)[160]<br>(c) 5.5 ppb (New York, USA)[161]<br>(d) 8.2 ppb (Salzburg, Austria)[162]<br>(e) 9 ppb (Munich, Germany)[162] | Aggriculture, vehicular exhaust | (a) 0.014 ppb MCC-IMS[163]<br>(b) 50 ppm AIMS[164]<br>(c) 18 ppb chemiresistor ($H_2SO_4$ solution)[165]<br>(d) 50 ppb chemiresistor ($MoO_3$)[166] | A |
| Acetone<br>67-64-1 | 435 | (a) 18.6 ppb (Ottawa, Canada)[167]<br>(b) 5.4 ppb (Melbourne, Australia)[168]<br>(c) 1.1 ppb (Sao Paulo, Brasil)[169]<br>(d) 2.4 ppb (Quinzhou, China)[170]<br>(e) 13.5 ppb (avg from several towns)[171]<br>(f) 7 ppb (Beijing)[172]<br>(g) 1.45 ppb (Zurich, Switzerland)[173] | Solvents, oxidation of NMHCs | (a) 0.02 ppb MCC-IMS[163]<br>(b) 14 ppb AIMS [174]<br>(c) 500 ppb AIMS[164]<br>(d) 20 ppb Si:$WO_3$ chemiresistor [175]<br>(e) 120 ppb Pt-$WO_3$ chemiresistor[176]<br>(f) 130 ppb optical spectroscopy[177]<br>(g) 170 ppb CTL ($Mn_3O_4$)[178] | A |
| 2-Butanone<br>78-93-3 | 9.8 | (a) 0.9 ppb (Ottawa)[167]<br>(b) 0.76 ppb (Quinzhou)[170]<br>(c) 1.35 ppb (avg from several Towns) [171]<br>(d) 0.2 ppb (Zurich, Switzerland)[173]<br>(e) 0.5 ppb (Niterói City, Brasil)[179] | Industrial solvent | (a) 11 ppb AIMS [174] | B |
| 2-Pentanone<br>107-87-9 | 3.2 | | | (a) 8 ppb AIMS [174] | B |
| 5-Hepten-2-one, 6-methyl-<br>110-93-0 | 16.6 | (a) 0.36 ppb (Rome)[180]<br>(b) 0.65 ppb (Milan)[180] | plants | (a) 0.7 ppb MCC-IMS [181] | A |
| 3-Buten-2-one<br>78-94-4 | 13.2 | (a) 0.17 ppb (Hong Kong)[145]<br>(b) 0.43 ppb (Nashville)[182] | oxidation of isoprene, vehicle exhaust | | B |
| 2,3 Butanedione<br>431-03-8 | 20 | | Food, kitchen waste | | B |
| Acetaldehyde<br>75-07-0 | 98 | (a) 9.4 ppb (Sao Paulo)[169]<br>(b) 2.4-45 ppb (Rio de Janeiro)[154]<br>(c) 8.0 ppb (Quinzhou)[170]<br>(d) 5.7 ppb (Beijing)[172]<br>(e) 3.6 ppb (Niterói City, Brasil)[179] | Ethanol fuel combustion, oxidation of NMHCs | (a) 500 ppb AIMS[164]<br>(b) 110 ppb bio-sniffer [183]<br>(c) 0.15 ppb chemiresistor (ZnO)[184]<br>(d) 500 ppb CTL [185] | A |
| n-Propanal<br>123-38-6 | 15.4 | (a) 0.4 ppb (Sao Paulo)[169]<br>(b) 0.35 ppb (Quinzhou)[170]<br>(c) 0.12 ppb (Zurich, Switzerland)[173]<br>(d) 0.83 ppb (Niterói City, Brasil)[179] | Disinfectant, kitchen waste, vehicle exhaust | (a) 25 ppb AIMS [174]<br>(b) 0.15 ppb chemiresistor (ZnO)[184]<br>(c) 250 ppb CTL ($ZrO_2$)[186] | A |
| 2-Propenal<br>107-02-8 | 5.8 | (a) 0.6 ppb (Sao Paulo)[169]<br>(b) 0.09 ppb (Zurich, Switzerland)[173] | vehicle exhaust | (a) 50 ppb AIMS[164] | B |
| 2-Propenal, 2-methyl<br>78-85-3 | 3.2 | (a) 0.1 ppb (Hong Kong)[145]<br>(b) 0.24 ppb (Nashville)[182]<br>(c) 0.02 ppb (Zurich, Switzerland)[173] | oxidation of isoprene, vehicle exhaust | | B |
| Propanal, 2-methyl-<br>78-84-2 | 1.3 | | kitchen waste, vehicle exhaust | | C |
| Butanal, 2-methyl-<br>96-17-3 | 1.5 | | kitchen waste | | C |
| Butanal, 3-methyl-<br>590-86-3 | 1.7 | | kitchen waste | | C |
| n-Hexanal<br>66-25-1 | 26 | (a) 1.2 ppb (Melbourne)[168]<br>(b) 0.35 ppb (Rome)[180] | Fuel combustion | (a) 24 ppb AIMS [174]<br>(b) 0.3 ppb MCC-IMS [13] | A |
| n-Heptanal<br>111-71-7 | 7.1 | (a) 0.4 ppb (Rome)[180] | Fuel combustion, atmospheric photooxidation of HCs, kitchen waste | | A |
| n-Octanal<br>124-13-0 | 8.2 | (a) 0.48 ppb (Rome)[180] | Fuel combustion, atmospheric photooxidation of HCs, kitchen waste | (a) 28 ppb AIMS [174]<br>(b) 0.1 ppb MCC-IMS [13] | A |
| n-Nonanal<br>124-19-6 | 13.6 | (a) 1.4 ppb (Melbourne)[168]<br>(b) 1.2 ppb (avg from several Towns) [171]<br>(c) 0.37 ppb (Rome)[180]<br>(d) 0.14 ppb (Niterói City, Brasil)[179] | Fuel combustion, atmospheric photooxidation of HCs, kitchen waste | (a) 0.3 ppb MCC-IMS [13] | A |
| Isoprene<br>78-79-5 | 36 | (a) 0.3 ppb (Seul)[93]<br>(b) 0.8 ppb (Karachi)[90]<br>(c) 0.13 ppb (Lille)[187]<br>(d) 0.34 ppb (Rome)[91]<br>(e) 0.252 ppb (Hong Kong)[145]<br>(f) 0.66 ppb (Guangzhou)[188]<br>(g) 0.41 ppb (Nashville)[182]<br>(h) 0.27 ppb (A Coruna)[146] | Plants, vehicular emissions | (a) 0.003 ppb MCC-IMS[163]<br>(b) 36 ppb MIR[189] | A |
| 1,3-Pentadiene, 2-methyl-, Z-<br>2787-45-3 | 0.3 | | | | C |
| 1,3-Pentadiene, 2-methyl-, E-<br>926-54-5 | 0.2 | | | | C |
| 2-Pentene, 2-methyl-<br>625-27-4 | 1.7 | | | | C |
| 1-Heptene<br>592-76-7 | 0.23 | | | | C |
| n-Heptane<br>142-82-5 | 0.5 | (a) 0.5 ppb (Seul)[93]<br>(b) 3.9 ppb (Karachi)[90]<br>(c) 9 ppb (Rio de Janeiro)[190]<br>(d) 0.05 ppb (Rome)[91]<br>(e) 0.56 ppb (Guangzhou)[188]<br>(f) 0.34 ppb (A Coruna)[146]<br>(f) 0.09 0.53 ppm (London) [155] | petrol evaporation, vehicle exhaust | | C |
| 1-Octene<br>111-66-0 | 0.5 | | | | C |
| n-Octane<br>111-65-9 | 1.5 | (a) 0.3 ppb (Seul)[93]<br>(b) 1.1 ppb (Karachi)[90]<br>(c) 1.2 ppb (Rio de Janeiro)[190]<br>(d) 0.1 ppb (Lille)[187]<br>(e) 0.79 ppb (Guangzhou)[188]<br>(f) 0.3 ppb (A Coruna)[146]<br>(g) 0.04 ppb (London) [155] | petrol evaporation, vehicle exhaust | | C |
| 1-Nonene<br>124-11-8 | 0.57 | | | | C |
| n-Nonane<br>111-84-2 | 2.5 | (a) 0.6 ppb (Seul)[93]<br>(b) 0.7 ppb (Karachi)[90]<br>(c) 2.1 ppb (Rio de Janeiro)[190]<br>(d) 0.95 ppb (avg from several Towns) [171] | petrol evaporation, vehicle exhaust | | C |
| Methanol<br>67-56-1 | 160 | (a) 8 ppb (Barcelona)[191]<br>(b) 22 ppb (avg from several Towns) [171]<br>(c) 14 ppb (Rio de Janeiro, Brasil)[97]<br>(d) 5.8 ppb (Osaka, Japan)[192]<br>(e) 1.8 ppb (Zurich, Switzerland)[173] | Solvents, biofuel evaporation, | (a) 100 ppm AIMS[164]<br>(b) 380 ppb CTL (nano-CdS)[193] | B |
| Ethanol<br>64-17-5 | 105 | (a) 37 ppb (Melbourne)[168]<br>(b) 64 ppb (avg from 50 studies) [171]<br>(c) 66.4 ppb (Rio de Janeiro, Brasil)[97]<br>(d) 8.2 ppb (Osaka, Japan)[192]<br>(e) 176.3 ppb (Sao Paulo, Brasil)[192]<br>(f) 6.6 ppb (Zurich, Switzerland)[173] | Solvents, biofuel evaporation, | (a) 0.525 ppb MCC-IMS[163]<br>(b) 55 ppb AIMS [174]<br>(c) 300 ppb biochemical[194]<br>(d) 700 ppb CTL [195] | C |
| 2-Propanol<br>67-63-0 | 69.6 | (a) 7.4 ppb (Ottawa, Canada)[167]<br>(b) 7.2 ppb (Osaka, Japan)[192]<br>(c) 44.2 ppb (Sao Paulo, Brasil)[192] | Disinfectants, antifreezers, biofuel evaporation | | C |
| Dimethylsulfide<br>75-18-3 | 9.0 | (a) 0.37 ppb (Seul, Korea) [196] | | | B |
| Allyl methyl sulfide<br>10152-76-8 | 0.5 | | | | C |
| Methyl propyl sulfide<br>3877-15-4 | 1.5 | | | | C |
| p-Cymene<br>99-87-6 | 0.8 | | Oxidation of α-pinene | | C |
| DL-Limonene<br>138-86-3 | 6.7 | (a) 9.4 ppb (Rio de Janeiro)[190]<br>(b) 19.6 ppb (Melbourne)[168]<br>(c) 3.7 ppb (avg from several Towns) [171] | Plants, Wood emission, food, kitchen waste | (a) 0.9 MCC-IMS[197] | C |
| α-Pinene<br>80-56-8 | 0.65 | (a) 2.4 ppb (Rio de Janeiro)[190]<br>(b) 6.2 ppb (Melbourne)[168]<br>(c) 0.2 ppb (Rome)[180]<br>(d) 0.21 ppb (Milan)[180]<br>(e) 0.19 (A Coruna)[146] | Plants, Wood emission, food, kitchen waste | (a) 0.9 MCC-IMS[197] | C |
| Furan, 2-methyl-<br>534-22-5 | 6.8 | | | | B |
| Furan, 3-methyl-<br>930-27-8 | 0.3 | (a) 0.05 ppb (USA)[198] | oxidation of isoprene, vehicle exhaust | | C |
| Furan, 2-pentyl-<br>3777-69-3 | 0.23 | | | | C |
| Acetonitrile<br>75-05-8 | 8.3 | (a) 0.12 ppb (Sydney)[199] | biomass burning | | B |
| γ-Butyrolactone<br>96-48-0 | 5.8 | | solvents | | B |
| Methyl acetate<br>79-20-9 | 1.6 | (a) 0.06 ppb (Zurich, Switzerland)[173] | Solvents, oxidation of MTBE and TAME | | B |
| Dimethyl selenide<br>593-79-3 | 0.14 | | | | C |

**Figures**

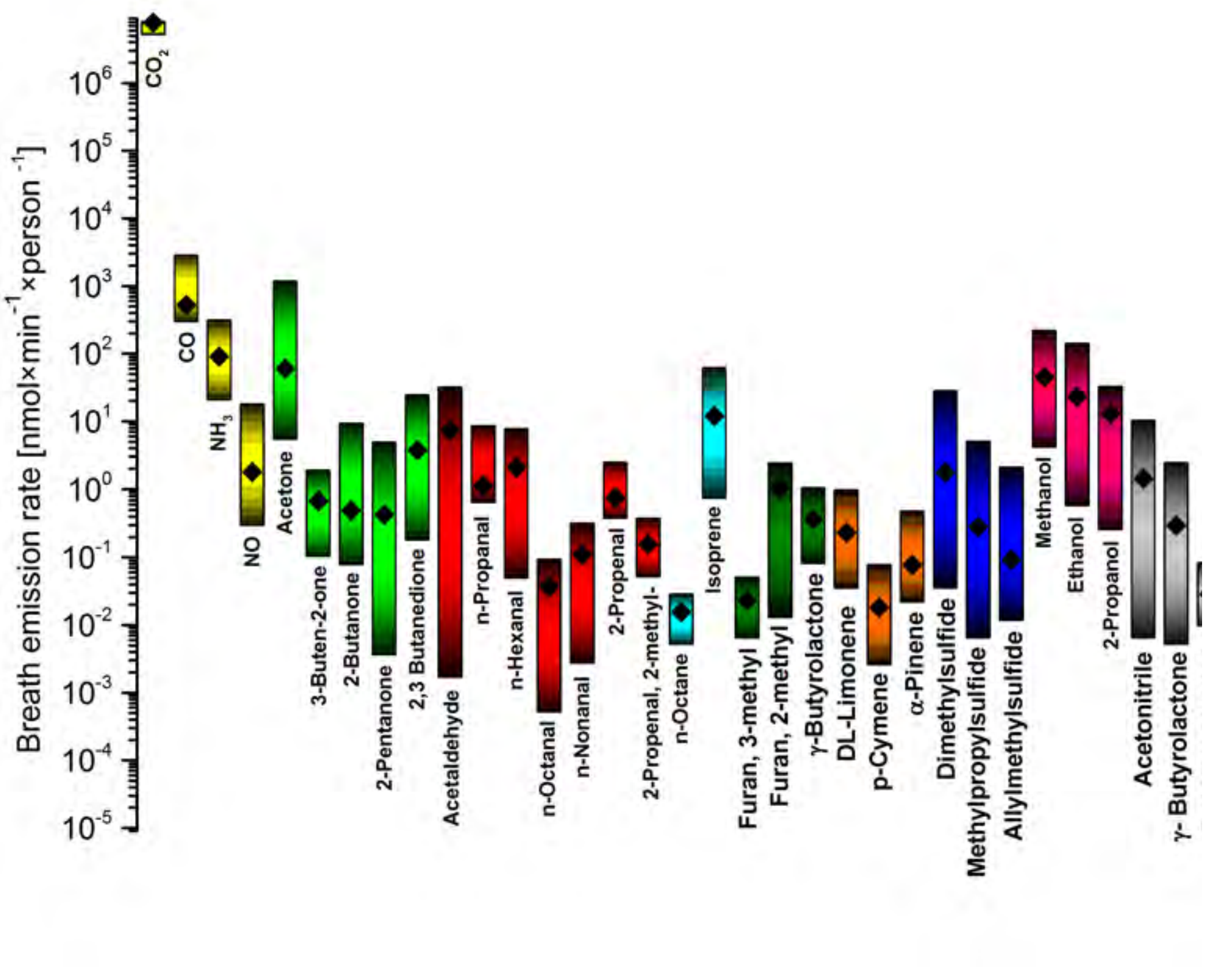

Figure 1 Ranges and means of emission rates of potential breath markers of human p

human body. Different colors correspond to the different chemical classes of compo calculated on the basis of reports indicated in column B of Table 1.

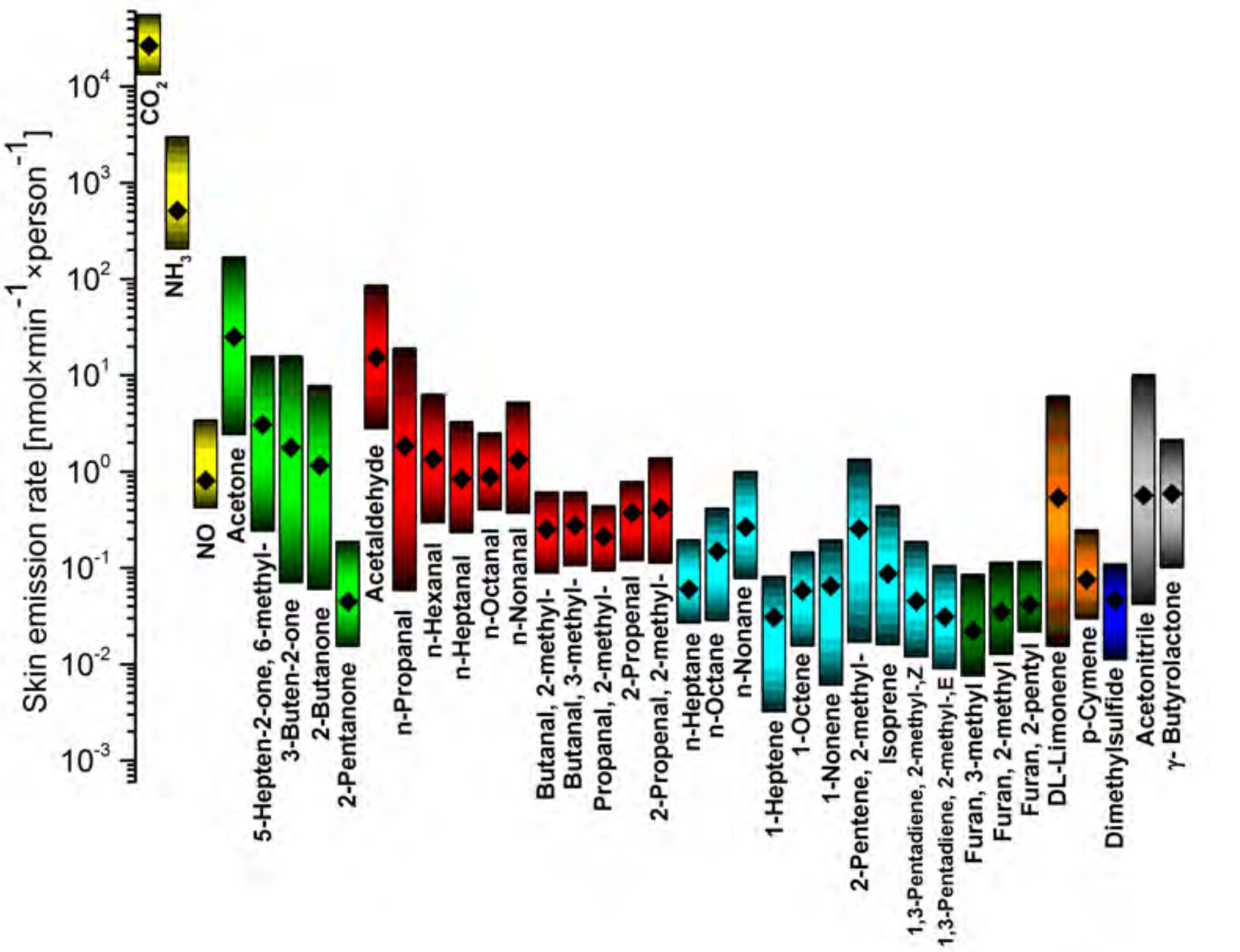

Figure 2 Ranges and means of emission rates of potential skin-borne markers of human presence from human body. The colors correspond to the different chemical classes of compounds. Ranges have been calculated on the basis of reports indicated in column C of Table 1.

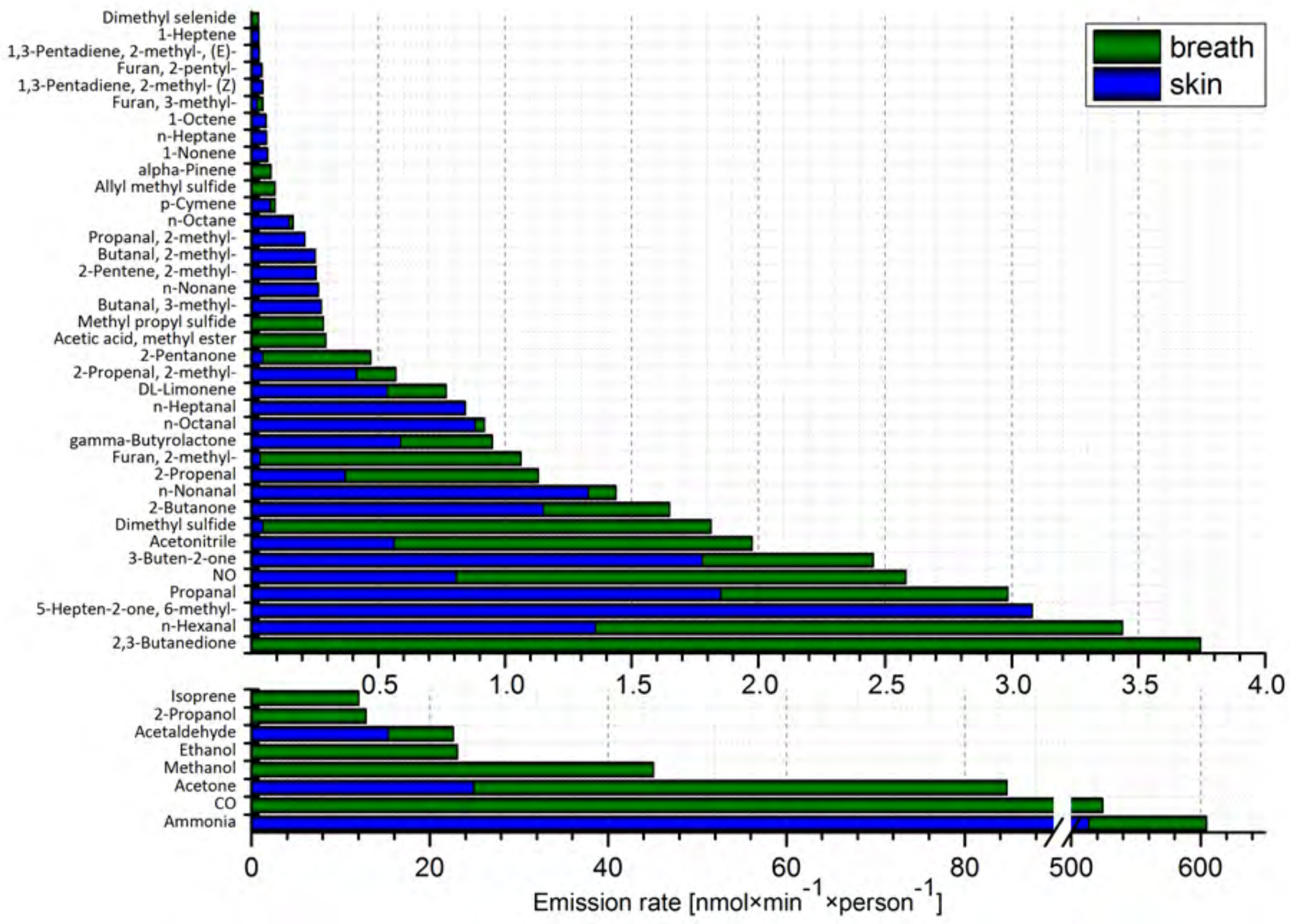

Figure 3 Total emission rates (considering means) of potential volatile markers of human presence from human body. $CO_2$ was excluded for clarity reasons.

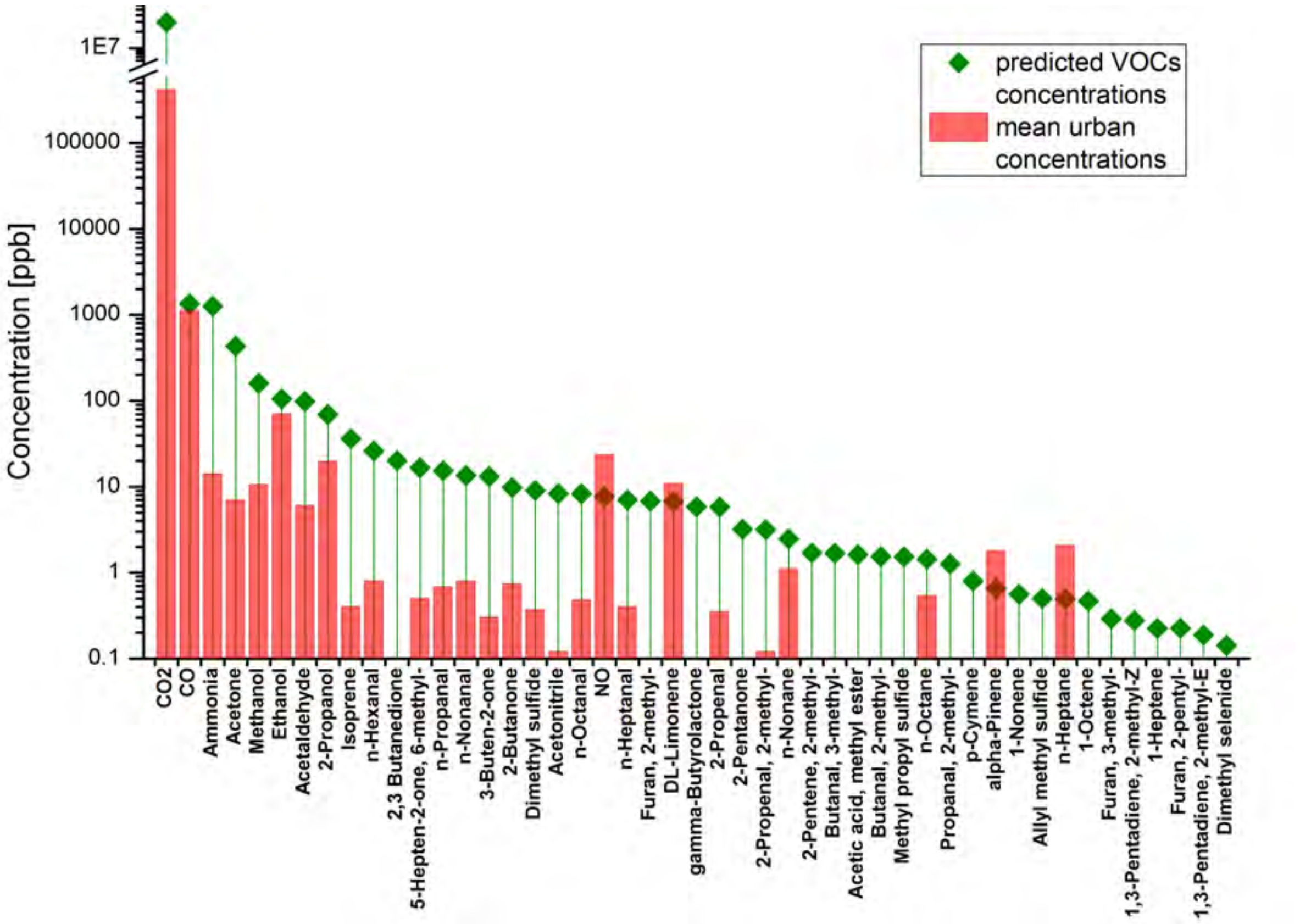

Figure 4 Exemplary chemical signature of entrapped person predicted for a point located 3m from a survivor and debris to air ratio 3:1. Red bars indicate mean urban air levels of compounds of interest.